\pdfoutput=1 

\documentclass{cidr-2019}

\usepackage{amsmath}
\usepackage{mathtools}
\usepackage{hyperref}             
\usepackage{listings}
\usepackage{color}
\usepackage{graphicx}
\usepackage{arydshln}
\usepackage{tikz-cd}
\usepackage{microtype}
\usepackage{paralist}

\newtheorem{theorem}{Theorem}[section]

\newcommand{\m}[1]{\ensuremath{\mathit{#1}}}
\newcommand\deltaAdd[1]{\Delta^{+}_{#1}}
\newcommand\deltaDel[1]{\Delta^{-}_{#1}}

\begin{document}


\title{Making View Update Strategies Programmable\\
--- Toward Controlling and Sharing Distributed Data ---}



%
%
%
%

\numberofauthors{1} 

\author{
%
%
\alignauthor
Yasuhito~Asano$^1$,
Soichiro~Hidaka$^2$,
Zhenjiang~Hu$^3$,
Yasunori~Ishihara$^4$,
Hiroyuki~Kato$^3$,
Hsiang-Shang~Ko$^3$,
Keisuke~Nakano$^5$,
Makoto~Onizuka$^6$,
Yuya~Sasaki$^6$,
Toshiyuki~Shimizu$^1$,
Van-Dang~Tran$^3$,
Kanae~Tsushima$^3$,
Masatoshi~Yoshikawa$^1$\\
\smallskip
       \affaddr{$^1$Kyoto University, $^2$Hosei University, $^3$National Institute of Informatics}\\
       \affaddr{$^4$Nanzan University, $^5$Tohoku University, $^6$Osaka University }\\
}

\maketitle

\begin{abstract}
Views are known mechanisms for controlling access of data and for sharing data of different schemas.
Despite long and intensive research on views in both the database community and the programming language community, we are facing difficulties to use views in practice. The main reason is that we lack ways to directly describe view update strategies to deal with the inherent ambiguity of view updating.

This paper aims to provide a new language-based approach to controlling and sharing distributed data based on views, and establish a software foundation for systematic construction of such data management systems. Our key observation
is that
\emph{a view should be defined through a view update strategy rather than a view definition}.
%
We show that Datalog can be used for specifying view update strategies whose unique view definition can be automatically derived, present a novel P2P-based programmable architecture for distributed data management where updatable views are fully utilized for controlling and sharing distributed data, and demonstrate its usefulness through the development of a privacy-preserving ride-sharing alliance system.
\end{abstract}


\section{Introduction}

Along with the continuous evolution of data management systems for the new market requirements, we are moving from centralized systems, which had often led to huge and monolithic databases, towards distributed systems, where data are maintained in different sites with autonomous storage and computation capabilities. The owner of the data stored on a site may wish to {\em control} and {\em share} data by deciding what information should be exposed and how its information should be used and updated by other systems.
This paper aims to provide a new language-based approach to controlling and sharing distributed data based on views, and establish a software foundation for systematic construction of such data management systems.

\subsubsection*{The view update problem in DB}


Views play an important role in controlling access of data \cite{FeSW81,Foster:2009:USV:1602936.1603620} and for sharing data of different schemas \cite{DoHI12,GHMT18}, since they were first introduced by Codd about four decades ago \cite{Codd74}.
A view is a relation derived from base relations, which is helpful to describe dependencies between relations and achieves database security.

Deeply associated with views is the classic {\em view update problem} \cite{Bancilhon:81,Dayal:82}: given a view definition in the form of a query over base relations, the view update problem studies how to translate updates made to the view to updates to the original base relations.
Despite a long and intensive study of view updating in the database community \cite{Bancilhon:81,Dayal:82,Keller:1986,Masunaga:2017:IAU:3022227.3022239}, there is no practical system that can fully support view updating.
This is because there are potentially many incomparable strategies (i.e., {\em ambiguity} of {\em view update strategies}) to translate updates to the base relations
 (e.g., deletion vs.~attribute value change for translating deletion when the view definition is a selection)%
, and it is difficult to choose a suitable one automatically \cite{Keller:1986}.

This calls for a general method to solve the fundamental tension
between expressiveness and realizability in the view update problem.
The richer language we use for defining views, the more difficult it becomes to find a suitable view update strategy.

\subsubsection*{Bidirectional transformations (BX) in PL}

To deal with this tension, researchers in the programming language community have generalized the view update problem to a general synchronization problem, and designed various domain specific languages  to support so-called {\em bidirectional transformation}s \cite{Lenses,Bohannon:08,Hidaka:10}.

A bidirectional transformation (BX) consists of a pair of
transformations: a forward and a backward transformation.
The {\em forward} transformation $\m{get}(s)$ accepts a source $s$ (which is a collection of base relations in the setting of view updating), and
produces a target view $v$, while the {\em putback} (backward)
transformation $\m{put}(s,v)$ accepts the original source $s$ and an updated view $v$,
and produces an updated source. 
These two transformations should be {\em
well-behaved} in the sense that they satisfy the following
round-tripping laws.
\[
\begin{array}{lllr}
\m{put}(s, \m{get}(s))  &=& s \qquad & \textsc{GetPut} \\
\m{get}(\m{put}(s, v)) &=& v & \textsc{PutGet}
\end{array}
\]
The \textsc{GetPut} property (or Acceptability~\cite{Bancilhon:81}) requires that no change on the view should
result in no change on the source, while the \textsc{PutGet}
property (or Consistency~\cite{Bancilhon:81}) demands that all changes to the view be completely translated
to the source by stipulating that the updated view should be the one computed by
applying the forward transformation to the updated source.
The exact correspondence between the notion of well-behavedness in BX and the properties on view updates such as translation of those under a constant complement \cite{Bancilhon:81,Dayal:82} has been extensively studied \cite{BeSm03}.

It has been demonstrated \cite{Bohannon:06} that this language-based approach helps to solve the view update problem with
a bidirectional query language, in which every query can be interpreted as both
a view definition and an update strategy.
However, the existing solution is unsatisfactory, because the view update strategies are chosen at design time and hardwired into the language, and what users wish to express may well not be included in the set of strategies offered by the language.

\subsubsection*{Problem: lack of control over view update strategies}

The main difficulty in using views to control and share distributed data lies in the inherent ambiguity of view update strategies when given a view definition (or a forward transformation).
We lack effective ways of controlling the view update strategy (or the putback transformation); it would be awkward and counterintuitive, if at all possible, to obtain our intended view update strategy by changing the view definition that is under our control, when the view definition becomes complicated.

We have taken it for granted that a view should be defined by a query and that a sound and intended update strategy should be automatically derived even if it is known that automatic derivation of an intended update strategy is generally impossible \cite{Keller:1986}. Now it is time to consider seriously the following two fundamental questions: (1) Must views be defined by queries over the base relations? and (2) Must view update strategies be automatically derived?

\subsubsection*{Our vision: a programmable approach}

We aim to solve the above problem, and provide a new language-based approach to controlling and sharing distributed data based on the view.
Our answer to the above two questions is:
\begin{quote}
{\em A view should be defined through a view update strategy to the base relations rather than a query over them}.
\end{quote}
This new perspective is in sharp contrast to the traditional approaches,
and it also gives a direction for solving the problem: view update strategies should be programmable.


This vision stems from the recent work on the putback-based approach \cite{HuPS14,FiHP15,KoZH16,KoHu18} to bidirectional programming. The key point is that although there are many \m{put}s that can correspond to a given \m{get}, there is at most one \m{get} that can correspond to a given $put$, and such \m{get} can be derived from \m{put}. Rephrasing this in the setting of view updating:
\begin{quote}
  {\em
  while
there may be many view update strategies for a given view definition, there is a unique view definition (if it exists) that corresponds to a view update strategy, and this view definition can be derived.}
\end{quote}
This new perspective on views implies that we should have a language for describing view update strategies and treat the view definition as a by-product of the view update strategy.

In this paper, we show that Datalog can be used for specifying view update strategies whose unique view definition can be derived, and present a novel P2P-based architecture called Dejima for distributed data management where updatable views are fully utilized for controlling and sharing distributed data. Our main technical contributions can be summarized as follows.

\begin{itemize}
  \item We show in Section \ref{sec:language} for the first time that Datalog is a suitable and powerful language for specifying various view update strategies, present a novel algorithm for automatically deriving the unique view definition from a view update strategy, and explain how such updatable views can be efficiently implemented using the trigger mechanism in PostgreSQL.


  \item We propose in Section \ref{sec:architecture} Dejima, a new P2P-based programmable architecture for distributed data management, where updatable views are fully utilized as a key component for controlling and sharing distributed data. In this architecture, each peer has full control of its data, {\em autonomously} managing its view update strategies and interact with other peers, while using a simple data synchronization mechanism between different peers for sharing data through views.


\item We validate this new approach in Section \ref{sec:implementation} through the development of a privacy-preserving ride-sharing alliance system. Being simple, this example gives a good demonstration of the need for controlling and sharing decentralized data.

\end{itemize}

A prototype implementation is available online \cite{Proto}, where the implementation code in OCaml and the tests of all the examples in this paper can be found.

\section{Foundation: Putback-based BX}
\label{sec:background}
\label{sec:putbx}

Before explaining our new view-based programming architecture, we
briefly review the theoretic foundation of BX.
As mentioned in the introduction, much research \cite{Lenses,Bohannon:06,Bohannon:08,Hidaka:10} on BX has been devoted to the {\em get-based}
approach, allowing users to write a forward
transformation and deriving a suitable putback transformation.
While the get-based approach is user-friendly,
a \m{get} function may not be injective, so there may exist
many possible  functions that can be combined with it to form a
BX. The usual solution is to enrich a get-based language with some putback information, but it remains awkward to control the choice of \m{put} behavior through
the change of enriched \m{get} programs.
The need for better control of \m{put} behavior is what makes bidirectional
programming challenging in practice.

In contrast to the get-based approach, the putback-based approach allows users to write the backward
transformation \m{put} and derives a suitable \m{get} that can be
paired with \m{put} to form a BX if it exists.
Interestingly, while \m{get} usually loses information
when mapping from a source to a view, \m{put} must preserve information
when putting back from the view to the source, according to the
\textsc{PutGet} property.

The most important fact is that ``putback''
is the essence of bidirectional programming \cite{HuPS14}.
That is, for a \m{put}, there exists at most one \m{get}
that can form a well-behaved BX with it. This is in sharp contrast to get-based
bidirectional programming, where many \m{put}s may be paired with a \m{get}
to form a BX.

\begin{theorem}[Uniqueness of \m{get} \cite{HuPS14}]
\label{lemma:injective}
Given a \m{put} function, there exists at most
one \m{get} function that forms a well-behaved BX.
\end{theorem}


\newcommand{\ruleeq}{\coloneq}

\section{Coding View Update Strategies}
\label{sec:language}

A good language for programming/coding view update strategies (i.e., putback transformations) should meet two requirements. First, it should be expressive enough to describe intended view update strategies; second, and more importantly, the corresponding view definition can be automatically derived. In this section, we show for the first time that it is possible to use Datalog to describe view update strategies. This may be surprising as Datalog is a language for describing queries rather than updates. Moreover, we present a novel algorithm that can automatically derive view definitions from view update strategies.
We also explain how such updatable views can be implemented using the trigger mechanism in PostgreSQL.

\subsection{Coding View Updates in Datalog}

Recall that a putback transformation accepts the original source and an updated view, and returns an updated source. We may simplify it as accepting the source and an updated view but returning an update on the source (rather than an updated source), because we can obtain the updated source by applying this update to the original source. Now, if we can represent source updates as a pair of insertion and deletion relations, we can formulate a putback transformation as a query that produces these two  relations from source relations and (updated) view relations. It is this observation that makes it possible to use Datalog, a query language, for writing putback transformations.

We consider Datalog with stratified negation and dense-order constraints
(i.e., the interpreted predicates $=$ and $<$) \cite{ceri1989}. A Datalog program basically consists of a set of rules (and facts) of the general form
\[
L_0 \ruleeq L_1, \ldots, L_n.
\]
where each $L_i$ is a literal of the form $p_i(t_1,\ldots, t_k)$ where $p_i$ is a query predicate and $t_1$,\ldots, $t_k$ are terms. A term is either a constant or a variable. We shall use $\deltaAdd{p}$ and $\deltaDel{p}$ to denote the \emph{delta} relations whose tuples are to be inserted to and deleted from the relation $p$, respectively.

Let us see how we can use Datalog to describe intended view update strategies. As a simple example, consider two sources, $s_1(X)$ and $s_2(X)$, and an updated view $v(X)$ that is expected to be the union of the two sources. Since the view has been updated, we can describe the following strategy to translate view updates to source updates.
\begin{align}
\deltaDel{s_1}(X) &\ruleeq s_1(X), \neg v(X). \label{eq:d1}\\
\deltaDel{s_2}(X) &\ruleeq s_2(X), \neg v(X).\label{eq:d2}\\
\deltaAdd{s_1}(X) &\ruleeq v(X), \neg s_1(X), \neg s_2(X). \label{eq:i1}
\end{align}
It reads: if a tuple is in $s_1$ but not in $v$, it should be deleted from $s_1$ by putting it into the deletion relation of $s_1$ (Rule (\ref{eq:d1})); if a tuple is in $s_2$ but not in $v$, then it should be deleted from $s_2$ (Rule (\ref{eq:d2})); and if a tuple is in $v$ but in neither $s_1$ nor $s_2$ (i.e., the tuple is newly inserted to $v$), it should be inserted to $s_1$ (Rule (\ref{eq:i1})).

Two remarks are worth making here. First, the above defines just one view update strategy, and there are indeed many others that can be used when the view is intended as the union of the two sources. For instance, we may replace Rule (\ref{eq:i1}) with the following rule
\begin{align}
\deltaAdd{s_2}(X) &\ruleeq v(X), \neg s_1(X), \neg s_2(X). \label{eq:i2}
\end{align}
to insert the tuple to $s_2$ instead of $s_1$ when a tuple is newly inserted to $v$, or choose to use both Rules (\ref{eq:i1}) and (\ref{eq:i2}) to insert the tuple to both $s_1$ and $s_2$.
Second, Datalog with stratified negation and dense-order constraints is {\em expressive} for us to describe various view update strategies on relations; in fact, all the view update strategies discussed in some previous work \cite{Bancilhon:81,Dayal:82,Masunaga:2017:IAU:3022227.3022239} can be specified in this variant of Datalog.


\subsection{Deriving View Definitions}

Suppose that a given putback transformation is well-behaved in the sense that there exists a view definition that can be paired with it to form a bidirectional transformation. We will not go into the detail about checking the validity of a putback transformation (e.g., based on the sufficient and necessary condition of Hu et al.~\cite{HuPS14}); rather, we will show informally how to derive the view definition when the given putback transformation is well-behaved.


Our derivation of the view definition is based on the \textsc{GetPut} property as discussed in the introduction. If $get(s)$ defines the view $v$, then $put(s,v) = s$ should hold.
Obviously, the constraint $put(s,v) = s$ in a Datalog update program means that all \textit{delta} relations must be empty after evaluation. By solving this constraint, we can establish the functional relationship $v =get(s)$ between the source $s$ and the view $v$. For example, recall the view update strategy defined by Rules (\ref{eq:d1}--\ref{eq:i1}). Let $\bot$ denote the empty relation. If the delta relations are empty, we have
\begin{align}
\bot &\ruleeq s_1(X), \neg v(X). \label{eq:d11}\\
\bot &\ruleeq s_2(X), \neg v(X).\label{eq:d21}\\
\bot &\ruleeq v(X), \neg s_1(X), \neg s_2(X). \label{eq:i11}
\end{align}
Now according to the following swapping law:
\[
p \ruleeq q, \neg r ~\Leftrightarrow~ r \ruleeq q, \neg p
\]
we can move the negative occurrences of $v$ in Rules (\ref{eq:d11}--\ref{eq:d21}) from the body to the head, and obtain the following
\begin{align}
v(X) &\ruleeq s_1(X). \label{eq:vdef1}\\
v(X) &\ruleeq s_2(X). \label{eq:vdef2}
\end{align}
where $\neg \bot$ is always true and omitted in the body. This is exactly what our view definition function is. It is worth noting that Rule (\ref{eq:i11}) is satisfied when $v$ is defined as above, which is easy to check; in fact, this is always true if the view update strategy is well-behaved.


\subsection{Implementing Updatable Views}

Given a view update strategy in Datalog,
once we obtain the corresponding view definition in Datalog,
we can implement a view that can be queried and updated.
Following Herrmann et al.'s approach \cite{DBLP:conf/sigmod/HerrmannVBRL17}, such a view is realized by automatically generating an ordinary view and a trigger in PostgreSQL.
The idea is to first generate equivalent SQL programs from the Datalog programs for the view definition and the view update strategy, and then encapsulate them into a view definition and a trigger definition.

To be concrete, recall the example in this section.
From the view definition in Rules (\ref{eq:vdef1}--\ref{eq:vdef2}),
our system can generate the following view definition in PostgreSQL:
  \lstdefinelanguage{SQL}{morekeywords={CREATE, TABLE, AS, SELECT, FROM, UNION, VIEW, OR, REPLACE, INSTEAD, OF, INSERT, DELETE, UPDATE, FOR, EACH, ROW, EXECUTE, PROCEDURE,
  ON, TRIGGER }}
  \lstset{basicstyle=\scriptsize,basewidth=0.7em,keywordstyle=\bfseries,escapechar=\%}
    \begin{lstlisting}[language=SQL]
CREATE OR REPLACE VIEW v AS
   SELECT a FROM s1
   UNION
   SELECT a FROM s2
  \end{lstlisting}
where {\tt a} is the attribute over the tables {\tt
  s1}, {\tt s2}, and {\tt v}.

Meanwhile, to support updates on the view {\tt v}, our system generates a
trigger named {\tt v\_trigger} defined by:
\lstset{basicstyle=\scriptsize,basewidth=0.7em,keywordstyle=\bfseries,escapechar=\%}
  \begin{lstlisting}[language=SQL]
CREATE TRIGGER v_trigger
   INSTEAD OF INSERT OR DELETE ON v
   FOR EACH ROW
   EXECUTE PROCEDURE v_proc();
\end{lstlisting}
which calls the procedure {\tt v\_proc()} every time a row is to be inserted to or
to be deleted from the view.
The procedure {\tt v\_proc()} realizes the view update strategy of Rules
(\ref{eq:d1}--\ref{eq:i1}).

\section{The Dejima Architecture}
\label{sec:architecture}


In Section~\ref{sec:language},
we have obtained a sophisticated mechanism
for making view update strategies programmable.
Toward controlling and sharing \emph{distributed} data,
we need one more mechanism, i.e., data synchronization.
In this section, 
we propose a P2P-based programmable architecture,
called the \emph{Dejima}\footnote{%
Dejima was the name of a small, artificial island located in Nagasaki, Japan.
All the trades between Japan and foreign countries were made through Dejima
from the middle of the 17th to the middle of the 19th century.
We use this name because the functionality is similar to Dejima.
}
architecture, where
each peer autonomously manages its view update strategy
and collaborates with other peers
through views called \emph{Dejima} tables.
Only a simple data synchronization mechanism
is necessary for the Dejima tables between peers.
Section~\ref{sec:p2parchitecture} gives
the definition of the Dejima architecture.
Section~\ref{sec:Comparison to Existing Architectures} highlights the
features of the Dejima architecture by comparing it 
with existing architectures, where
controlling and sharing with update are limited or not supported.

\subsection{Definition of the Dejima Architecture}
\label{sec:p2parchitecture}

\begin{figure}[t]
  \centerline{\includegraphics[scale=.148,clip,trim=60mm 30mm 20mm 10mm]{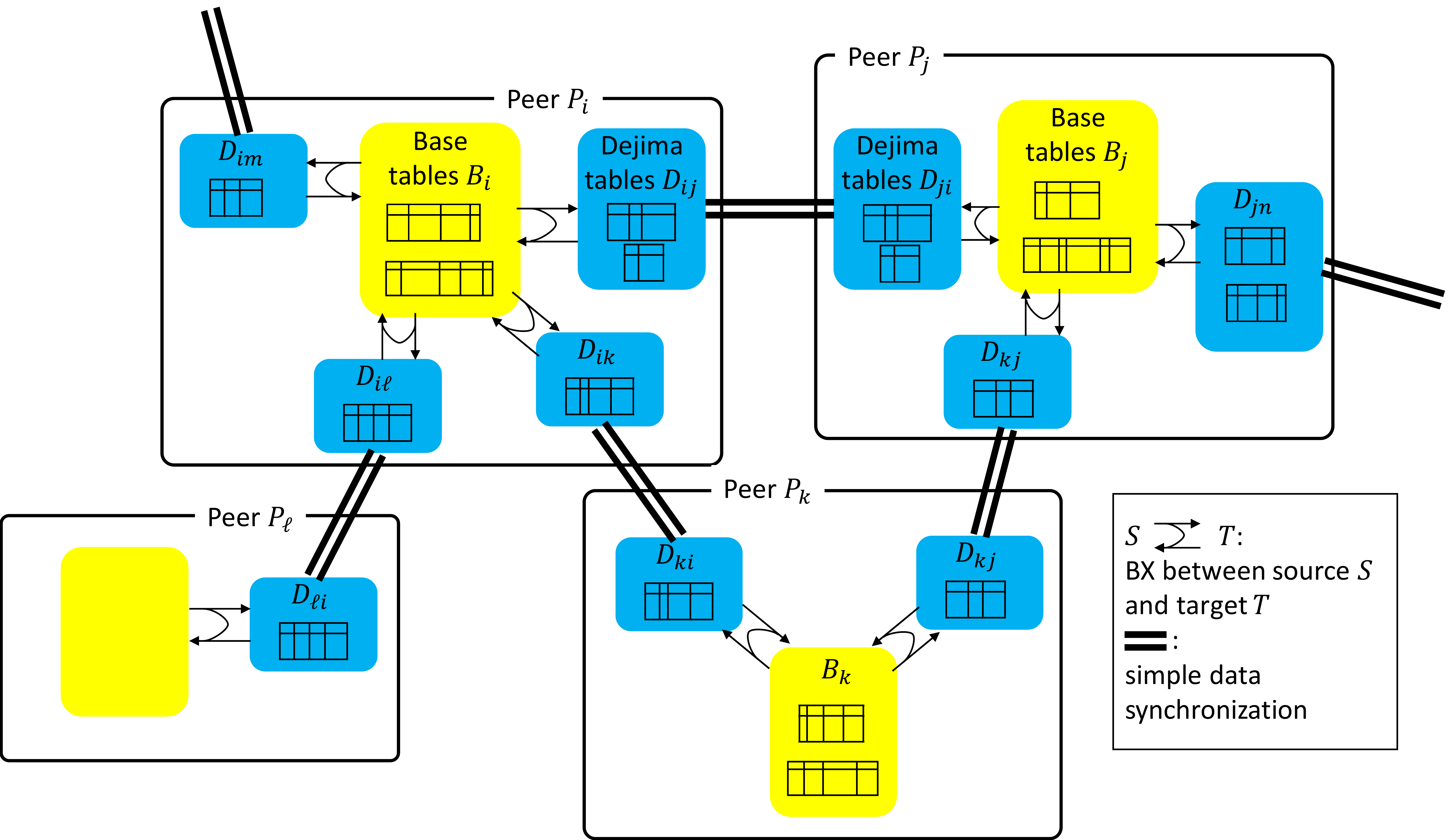}}
    \vspace{-0.2cm}
  \caption{The Dejima Architecture}
  \label{fig:p2p_architecture}
\end{figure}

Suppose that two peers $P_i$ and $P_j$ agree to share subsets of
their own data $B_i$ and $B_j$, respectively.
In the Dejima architecture (Figure~\ref{fig:p2p_architecture}),
peer $P_i$ prepares a set $D_{ij}$ of Dejima tables,
which is a set of views of $B_i$ to be shared with $P_j$
(and similarly, $P_j$ prepares $D_{ji}$).
Here is a novel concept of data sharing in the Dejima architecture:
$P_i$ and $P_j$ achieve data sharing between them
by maintaining $B_i$ and $B_j$ so that the equation $D_{ij}=D_{ji}$ holds.

To be specific, $P_i$ prepares $D_{ij}$
by specifying view update strategies for BX between $B_i$ and $D_{ij}$.
The BX represents
what information of $P_i$ should be exposed to $P_j$
and how the information of $P_i$ should be updated by $P_j$.
$D_{ij}$ and $D_{ji}$ must have the same schema.
Then, $P_i$ and $P_j$ continue to update $B_i$ and $B_j$, respectively,
so that the equation $D_{ij}=D_{ji}$ holds,
according to their view update strategies.

Now, we describe the Dejima architecture formally.
Let $P_1$, \ldots, $P_n$ be participating peers, where
each $P_i$ has a set $B_i$ of its base tables, that is,
original tables owned by $P_i$.
Let $D_{ij}$ be the set of Dejima tables from $P_i$ to $P_j$,
where $D_{ij}$ and $D_{ji}$ have the same schema.
Let $\m{get}_{ij}$ and $\m{put}_{ij}$ be the get and put functions
between $B_i$ and $D_{ij}$.
Then, the following equations hold in our architecture:
\begin{eqnarray*}
\m{get}_{ij}(B_i) & = & D_{ij},
\\
\m{put}_{ij}(B_i',D_{ij}) & = & B_i,
\\
D_{ij} & = & D_{ji}.
\end{eqnarray*}
In the second equation,
$B_i'$ denotes the ``previous version'' of $B_i$ and
$\m{get}_{ij}(B_i')$ has been just updated to $D_{ij}$.

In general,
$D_{ij}$ derived from $B_i$ and
$D_{ji}$ derived from $B_j$ are different
just after $P_i$ and $P_j$ agreed to share them.
We need some protocol for such initial synchronization of
$D_{ij}$ and $D_{ji}$, depending on the application.
One of the simplest protocols would be as follows:
the initiator, say $P_i$, of the agreement changes
the set $B_i$ of its base tables so that $D_{ij}$ becomes equal to $D_{ji}$.
Interestingly,
once $D_{ij}$ and $D_{ji}$ are synchronized,
they need not be materialized.
To see this, suppose that $P_i$ has just made some update
$\Delta B_i$ to $B_i$,
where $\Delta B_i$ involves insertion, deletion, and modification.
Using some operators informally,
the update $\Delta D_{ij}$ to be made to $D_{ij}$ would be represented as:
\[
\Delta D_{ij} = \m{get}_{ij}(B_i+\Delta B_i) - \m{get}_{ij}(B_i),
\]
which can be computed from $B_i$ and $\Delta B_i$.
Moreover, since $D_{ij}=D_{ji}$, the update $\Delta D_{ji}$
to be made to $D_{ji}$ must be equal to $\Delta D_{ij}$,
which is sent from $P_i$ to $P_j$
through the synchronization mechanism.
Hence, the update $\Delta B_j$ to be made to $B_j$ is:
\begin{eqnarray*}
\Delta B_j &=& \m{put}_{ji}(D_{ji}+\Delta D_{ji},B_j) - B_j
\\
&=& \m{put}_{ji}(\m{get}_{ji}(B_j) + \Delta D_{ij}, B_j) - B_j,
\end{eqnarray*}
which can be computed from $B_j$ and $\Delta D_{ij}$.

Note that update $\Delta B_i$ is propagated to peers indirectly
connected to $P_i$.
Also note that one of the peers directly or indirectly connected to
$P_i$ may reject the propagated update because of
its view update strategy.
For such cases, recovery is made by undoing $\Delta B_i$.
Hence, we need to leverage ACID or BASE transaction.
It is our future work how to manage transactions and
how to control the cascading update propagation
in the P2P-based Dejima architecture.



\subsection{Comparison to Existing Architectures}
\label{sec:Comparison to Existing Architectures}

Piazza~\cite{Halevy2003,Halevy2004} is one of the first projects on
peer-to-peer-based data sharing.
The Piazza system is for sharing distributed XML documents
without using global ontologies.
It provides query answering functionality based on the certain answer
semantics by rewriting given queries.
Updating XML documents on peers is not supported by Piazza.

\textsc{Orchestra}~\cite{Ives2005,Karvounarakis2013} is
a successor project of Piazza.
This project is motivated by the need for collaborative sharing of
scientific data.
The novel concept is referred to as
\emph{collaborative data sharing systems} (CDSS for short),
where data inconsistency between different peers is positively
allowed because of this motivation.
In CDSS, every peer can autonomously import a copy of other peers' data,
modify the imported copy,
merge the modified data with its original data,
and then publish the merged data to other peers.
Hence, write access to other peers is not allowed.
This feature will be a drawback when we implement a distributed
business data management system like a ride-sharing alliance system.



As mentioned by Arenas et al.~\cite{Arenas2003},
PeerDB~\cite{Ng2003} is the first implementation of
a peer-to-peer data sharing system.
Neighbor peers are loosely connected by schema matching rather than
schema mapping.
When a user issues a query to a peer,
PeerDB traverses over the connected peers and
identifies candidate relations for which the query is evaluated.
After the user selects appropriate relations,
the query is actually evaluated.
However, updating data on other peers does not seem to be supported
in PeerDB.
\section{Ride-Sharing Application}
\label{sec:implementation}

To explain our Dejima architecture and implementation concretely,
let us consider a simple example of ``privacy-preserving ride-sharing alliance system''.
This example gives a good demonstration of the need for controlling and sharing distributed data in the P2P-based Dejima architecture.

%
%
\subsection{Requirements and System Design}
Ride-sharing has become a popular application which allows non-professional drivers to provide taxi service using their vehicles. Each driver/vehicle usually belongs to a single ride-sharing company.
As the size of the ride-sharing market increases, it is expected that ``alliances'' will be formed among companies so that they can share the passengers' requests.

A system for a ride-sharing alliance receives requests from passengers and then identifies and books an appropriate vehicle for each request.
There are three major requirements for the system:
1) each company autonomously works on its own and collaborates with other companies in the same alliance,
2) privacy protection is indispensable; passengers might not disclose their precise locations to many drivers, and drivers belonging to a company might not want to disclose their precise locations to other companies, and
3) the collaboration (query and update) is made between the companies and passengers through the alliance.

\begin{figure}[t]
  \centerline{\includegraphics[scale=.38]{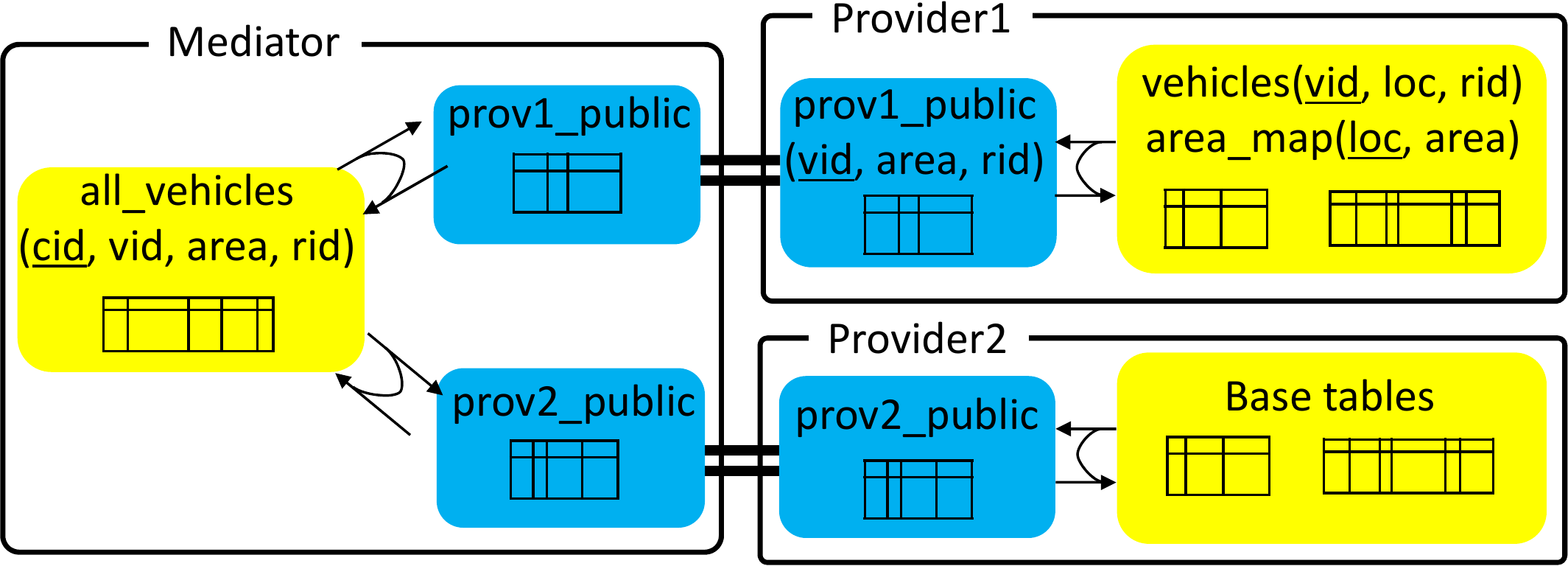}}
    \vspace{-0.3cm}
  \caption{System for ride-sharing alliance}
  \label{fig:ride-sharingsystem}
    \vspace{-0.3cm}
\end{figure}

Our Dejima architecture satisfies all the above requirements as shown in Figure \ref{fig:ride-sharingsystem}.
First, each of the companies (depicted as Provider in the figure) and alliances (Mediator) is implemented as a peer which works and collaborates with other peers.\footnote{It is our option for Dejima architecture to choose ACID transaction for tight collaboration between peers.}
Second, the companies protect the location privacy of the vehicles and passengers by using their own privacy policy.
Third is for the collaboration.
The mediators and providers collaborate with each other through views: the disclosed data at companies are propagated to the mediator through Dejima tables and the mediator recommends vehicles to the passengers.
They also collaborate through view update strategies: once a vehicle is assigned to a passenger (by the mediator), the update made on the local data of the mediator is propagated to the company the assigned vehicle belongs to.
Notice that the Dejima tables on the mediator is updatable without causing update ambiguities, which is achieved by our BX-based view update strategies.

\subsection{Updatable View Implementation}


Let us have a close look at Figure \ref{fig:ride-sharingsystem}, where we have three peers, a mediator and two ride-sharing companies (Provider 1 and Provider 2), and demonstrate in detail how to implement bidirectional transformations between the local data and the Dejima tables in each peer through description of view update strategies.

To be concrete, we assume that Provider 1 has its own vehicle data in two tables:
\begin{itemize}
  \item \texttt{vehicles(\underline{vid}, loc, rid)}, which contains for each vehicle its identifier, location, and the request identifier assigned to the vehicle, and
 \item \texttt{area\_map(\underline{loc}, area)}, which is used to obfuscate the precise locations of the vehicles by associating each location with a less precise area.
\end{itemize}
These two tables are managed in a private DBMS and allowed partial access from the mediator through a Dejima table:
\[
\texttt{prov1\_public(\underline{vid}, area, rid)}
\]
where, for privacy protection, only the area information (instead of the precise location information) of a vehicle is disclosed.
We assume, for simplicity, that Provider 2 is the same as Provider 1, although there is no problem for Provider 2 to have a completely different structure with different privacy policy on its local data.
Moreover, we assume that the mediator has a mediation table
\[
\texttt{all\_vehicles(\underline{cid}, \underline{vid}, area, rid)}
\]
which contains an additional column of company identifiers (\texttt{cid})
compared with the Dejima table \texttt{prov1\_public} used to synchronize with Provider 1.
%


To realize this riding-sharing alliance system, we basically need to write two view update strategies (since we assume that Provider 2 has the same structure as Provider 1). One is to propagate the update of the Dejima table \texttt{prov1\_public} to the mediation table \texttt{all\_vehicles} in the mediator. The other is to propagate the update of the Dejima table \texttt{prov1\_public} to the sources \texttt{vehicles} and \texttt{area\_map} for controlling local data access by Provider 1.

For the former, we may simply define the following view update strategy
\begin{equation*}
  \small
  \begin{array}{lll}
  \deltaDel{\m{all\_vehicles}} (C, V, A, R)  \ruleeq& \m{all\_vehicles} (C, V, A, R),  \\
     & C=1, \neg \m{prov1\_public} (V, A, R).\\
  \deltaAdd{\m{all\_vehicles}} (C, V, A, R)  \ruleeq& \m{prov1\_public} (V, A, R),  \\
   & C=1, \neg \m{all\_vehicles} (C, V, A, R).
  \end{array}
\end{equation*}
which reads: a tuple $(1, V, A, R)$ in \texttt{all\_vehicles} should be deleted if the tuple $(V, A, R)$ is not in \texttt{prov1\_public}, and should be inserted to \texttt{all\_vehicles} if it is not in \texttt{all\_vehicles} but the tuple $(V, A, R)$ is in \texttt{prov1\_public}.

For the latter, describing a view update strategy for the sources is more interesting because we want to describe the strategy where any change on the view is propagated only to \texttt{vehicles} while \texttt{area\_map} is kept unchanged.  This view update strategy is defined by
\begin{equation*}
  \begin{array}{lll}
  \deltaDel{\m{vehicles}} (V, L, R)  \ruleeq & \m{vehicles}(V, L, R), \m{area\_map}(L,A), \\
    & \neg \m{prov1\_public} (V, A, R). \\
  \deltaAdd{\m{vehicles}} (V, L, R)  \ruleeq & \m{prov1\_public} (V, \_, R), \\
  & \m{vehicles}(V, L, R'), R \neq R'.\\
  \end{array}
\end{equation*}
where we define the delta relations only for \texttt{vehicles} while using
\texttt{area\_map} as a reference relation. Note that in the context of ride-sharing,
the change on \texttt{prov1\_vehicles} is just the modification of the \texttt{rid} values of some
tuples for assigning vehicles. As in the above view update strategy, such modification is reflected to \texttt{vehicles} by deleting the tuples with the old \texttt{rid} values using the first rule and inserting the tuples with the new  \texttt{rid} values using the second rule.

This is all what one needs to write for implementing updatable views for change propagation. Now our tool can automatically derive the corresponding view definitions and generate the view definitions and triggers in PostgreSQL \cite{Proto}. We omit the details of the generation here, but for the two view update strategies defined for our ride-sharing alliance system, we can automatically derive the view definition for \texttt{prov1\_public} in the mediator:
\begin{equation*}
    \begin{array}{rl}
    \m{prov1\_public} (V, A, R) \ruleeq \m{all\_vehicles} (C, V, A, R), C=1.\\
\end{array}
\end{equation*}
and the view definition for \texttt{prov1\_public} in Provider 1:
\[
\m{prov1\_public} (V, A, R) \ruleeq \m{vehicles}(V,L,R), \m{area\_map}(L,A).
\]

\section{Conclusion}
\label{sec:conclusion}

In this paper, we have proposed a novel perspective on views, which are defined using view update strategies rather than queries over base relations.
This perspective stems from the studies of bidirectional transformations within the programming language community, in particular the insight that well-behaved queries are uniquely determined by, and can be derived from, view update strategies.
We have shown that updatable views play an important role in the design a programmable P2P-based architecture for controlling and sharing distributed data, and that these updatable views can be constructed through description of intended view update strategies in Datalog. We have implemented a prototype system and demonstrated its usefulness in the development of a  privacy-preserving ride-sharing alliance system.

We believe that it is worth reporting as early as possible the new perspective on views and the view-based programmable data management architecture arising from it, so that researchers in databases and programming languages can start working together to explore this promising direction.



\small
\bibliographystyle{abbrv}
\bibliography{references}  


%
%
%
%

\end{document}